\documentclass[aps,pra,reprint,longbibliography,amsmath,amssymb,multicol]{revtex4-2}
\usepackage[dvipdfmx]{graphicx}
\usepackage{dcolumn}
\usepackage{bm}
\usepackage{physics}
\usepackage{float}
\usepackage{color}
\usepackage{ulem} 
\usepackage{color}

\begin{document}

\preprint{APS/123-QED}

\title{Proposal for experimental realization of quantum spin chains with quasiperiodic interactions using Rydberg atoms}

\author{Takaharu Yoshida}
\email{1225707@ed.tus.ac.jp}
\author{Masaya Kunimi}%
\email{kunimi@rs.tus.ac.jp}
\author{Tetsuro Nikuni}
\email{nikuni@rs.tus.ac.jp}

\affiliation{%
 Department of Physics, Tokyo University of Science, Shinjuku, Tokyo, 162-8601, Japan
}%


\begin{abstract}
Investigating localization properties of interacting disordered systems plays a crucial role in understanding thermalization and its absence in closed quantum systems. However, simulating such systems on classical computers is challenging due to their complexity. In this work, we propose a method to realize $S=1/2$ and $S=1$ quantum spin models with quasiperiodic interactions using Rydberg atoms by utilizing the high tunability of the atoms' spatial positions. We also perform numerical calculations and show that these models host a many-body critical regime, which differs from the ergodic and many-body localized regimes.
\end{abstract}

\maketitle

\section{\label{sec:level1}Introduction}

Due to the recent experimental progress in quantum technology, research on thermalization in closed quantum systems has become one of the most active research areas over the decades. The eigenstate thermalization hypothesis (ETH)~\cite{Deutsch1991-vy,Srednicki1994-dl,Rigol2008-rf} is an important concept for explaining the thermalization mechanism. The ETH claims that as long as we consider local or few-body observables, the expectation value taken with the energy eigenstates of the Hamiltonian is equal to that taken with the microcanonical ensemble. When the system satisfies the strong version of the ETH, we can easily prove that the system will thermalize after sufficiently long unitary time evolution~\cite{DAlessio2016-fh,Mori_undated-rz}. Then, the ETH can be regarded as the quantum counterpart of the ergodic hypothesis. Although no mathematical proof exists for the strong ETH, many numerical and experimental results show that almost all nonintegrable systems satisfy the ETH and reach a thermal equilibrium state.

It is known that some systems do not satisfy the strong ETH. Integrable systems are  examples of violating the strong ETH~\cite{Rigol2007,Rigol2009,Rigol2009-2,Vidmar_2016}, and they relax to a nonthermal state whose expectation values of local observables are described by the generalized Gibbs ensemble ~\cite{Rigol2007,Vidmar_2016}. Another example is quantum many-body scar states ~\cite{Bernien2017-om,Turner2018-lc,Schecter2019-nk,Papic2021-ms,Moudgalya2022-vy,Chandran2023-qk}, which are special energy eigenstates of the system and have a small amount of entanglement entropy despite being highly excited states. A further example is Hilbert space fragmentation~\cite{Khemani2020,Sala2020,Moudgalya2021,Scherg2021-jg}, which occurs when some kinetic constraints, such as dipole conservation, are present.

Many-body localization (MBL) is another example of an ETH-violating state. This state appears when interacting systems have sufficiently strong random or quasiperiodic potentials~\cite{Huse2014-sh,Nandkishore2015-he,Schreiber2015-th,Abanin2019-ck}. The mechanism of the ETH violation in the MBL systems is the emergence of integrals of motion~\cite{Imbrie2016-cj}.  Although there are many studies on MBL, its existence in the thermodynamic limit is still controversial~\cite{Suntajs2020,Sels2021-ny,De_Tomasi2021-mj,Abanin2021-rm,Morningstar2022-kk}.

Recently, several papers have reported a new localization phase called the critical phase~\cite{Wang2020-ba,Wang2023-lt,Duncan2023-oy}. This phase can arise not only from quasiperiodic magnetic fields or potentials but also from quasiperiodic interactions or hoppings. Other studies have discussed the robustness of such critical regimes in many-body systems~\cite{Wang2021-zy,Zhou2023-ve}. The critical regime is considered to be different from the MBL regime because the fractal dimension  and level spacing statistics take different values. Although this novel localized state is interesting, its existence in the thermodynamic limit is still unclear due to the limitations on the system sizes accessible to numerical calculations.

To overcome these problems, one can use quantum simulations~\cite{Bloch2008-ug,Altman2021-AO}. In this work, we focus on the Rydberg-atom quantum simulators~\cite{Saffman2010-am,Browaeys2016-fu,Browaeys2020-eh,Morgado2021-ai}. The Rydberg atom (state) has a large principal quantum number $n\gg1$. Due to the large $n$, the Rydberg atom is highly polarizable. Consequently, strong dipole-dipole or van der Waals interactions exist between two Rydberg atoms. Combining the properties of the Rydberg atoms and the optical tweezer technique, one can realize highly controllable quantum spin systems in various lattice geometries. For example, the Rydberg-atom quantum simulators have realized the Ising, $XY$, and $XXZ$ models~\cite{Bernien2017-om,Orioli2018-Ro,Signoles2021-gd,Franz2024-Ob}.

This work proposes a method to realize spin-1/2 and spin-1 quantum spin chains with quasiperiodic interactions in the Rydberg-atom quantum simulator. Our method is based on the anisotropy of the dipole-dipole interaction and controllability of the position of the atoms owing to the optical tweezer techniques. In the spin-1/2 case, the situation is similar to that in Ref.~\cite{De_Leseleuc2019-xu}. In the spin-1 case, we use the F\"{o}rster resonance to realize the system with the dipole-dipole interaction ~\cite{Ravets2015-bz,Chew2022-ge}. We also study the nonergodic behavior of our proposed models by calculating the level spacing ratio, entanglement entropy, and dynamics.

The remainder of this paper is organized as follows. In Sec.~\ref{setup} we discuss the procedure to realize quasiperiodic interactions using Rydberg atoms. In Sec.~\ref{model} we explain our models. In Sec.~\ref{results}, we show numerical results for some observables. In Sec. \ref{Stat} we show the results of eigenstate statistics, confirming whether the system satisfies the ETH. In Sec.~\ref{Dyna} we show the time evolution of physical quantities. In Sec.~\ref{sum} we summarize our results and discuss long-time experiments. In the Appendixes, we discuss the technical details.

\section{\label{setup}Setup}

\begin{figure}[h]
\centering
\includegraphics[width=8.6cm,clip]{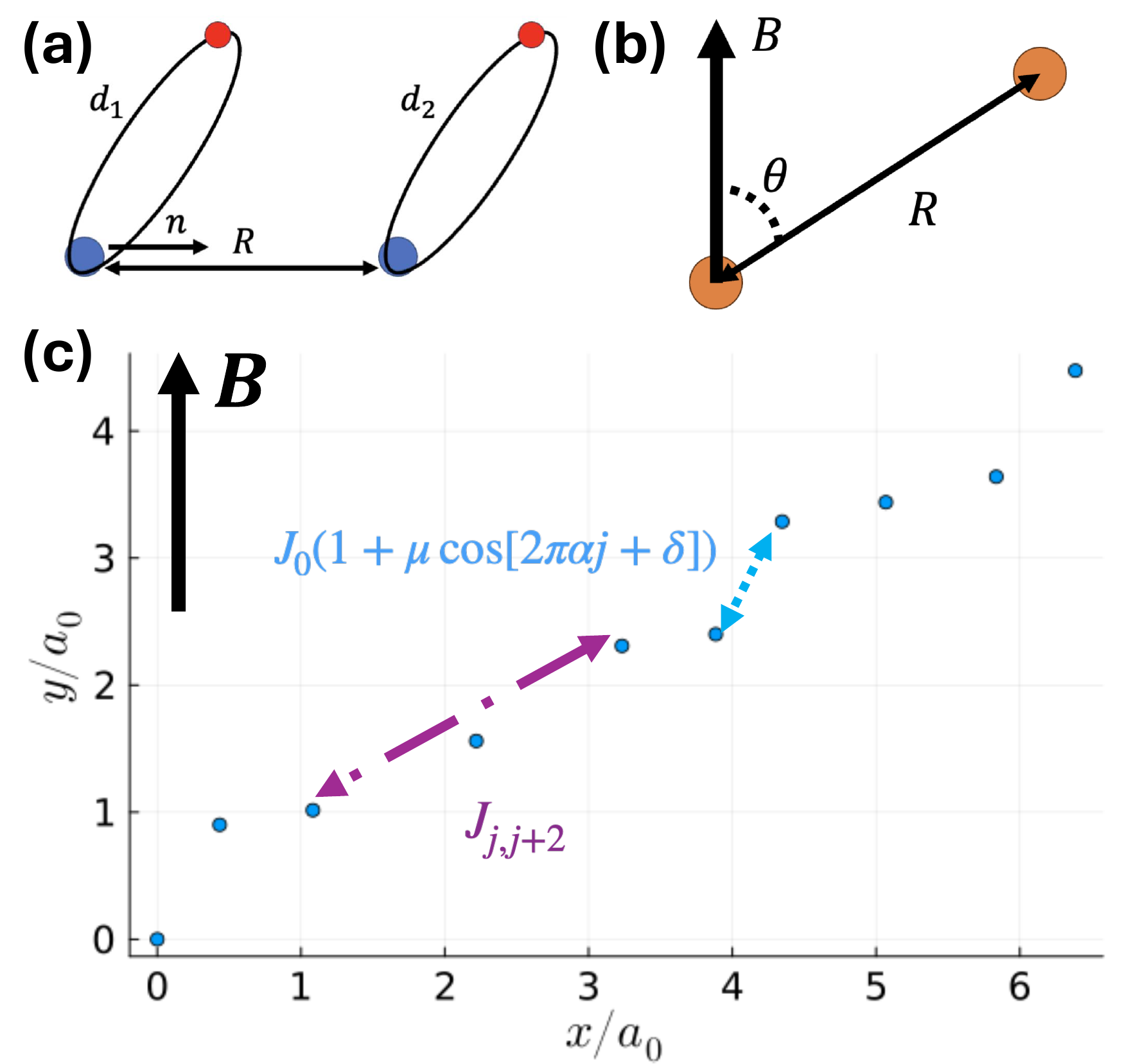}
\caption{Schematic of our procedure. (a) Schematic of two Rydberg atoms and their interaction. (b) Schematic of two atoms under a weak magnetic field. (c) Example of atom positions for $L=10$ and $\mu=2.5$. }
\label{schematicpic}
\end{figure}%

In this section, we show how to implement a spin chain with quasiperiodic interactions in the Rydberg-atom quantum simulator. First, we consider two Rydberg atoms labeled by 1 and 2 located in the $xy$ plane with distance $R$ along the direction ${\bm n}$, where $\bm{n}$ is a unit vector [see Fig.~\ref{schematicpic}(a)]. Because of the large polarizability, the dipole-dipole interaction acts between Rydberg atoms~\cite{Browaeys2016-fu,Weber2017-ji}:
\begin{equation}
    \hat{V}_{{\rm dd}}=\frac{1}{4\pi\varepsilon_0}\frac{\hat{\boldsymbol{d}}_1\cdot\hat{\boldsymbol{d}}_2-3(\hat{\boldsymbol{d}}_1\cdot \boldsymbol{n})(\hat{\boldsymbol{d}}_2\cdot \boldsymbol{n})}{R^3},
\label{ddint}
  \end{equation}
where $\varepsilon_0$ is the electric constant, and each electric dipole operator is denoted 
by $\hat{\bm{d}}_i\;(i=1,2)$.

Here, we discuss the spin-1/2 case. In this case, we assign $\ket{\uparrow}$ and $\ket{\downarrow}$ states as $\ket{nS}$ and $\ket{n'P}$ $(n\simeq n'\gg1)$ Rydberg states, respectively. Since the states $\ket{nS}$ and $\ket{n'P}$ have opposite parity, the dipole-dipole interaction appears at first order in perturbation theory. Owing to the selection rule, the nonvanishing term of the dipole-dipole interaction is the exchange process: $\ket{nS,n'P}\leftrightarrow\ket{n'P,nS}$. Using the spin language, we obtain the $XY$ Hamiltonian:
\begin{align}
\hat{H}&=\sum_{i<j}2J_{ij}(\hat{S}_i^x\hat{S}_j^x+\hat{S}_i^y\hat{S}_j^y),\label{eq:definition_of_S=1/2_Hamiltonian}\\
J_{ij}&\equiv \frac{C_3(\theta_{ij})}{R_{ij}^3}\equiv \frac{C_3(1-3\cos^2\theta_{ij})}{R_{ij}^3},\label{eq:definition_of_Jij}
\end{align}
where $\hat{S}_i^{\mu}\;(\mu=x,y,z)$ is the spin-1/2 operator at the $i$th site, $J_{ij}$ represents the interaction strength, $C_3$ is the interaction coefficient, $R_{ij}$ is the distance between the $i$th and $j$th atoms, $\theta_{ij}$ is the angle between the quantization axis and the atom pair [see Fig.~\ref{schematicpic}(b)]. Here, we define the quantization axis as the $y$-axis by assuming that the magnetic field is applied in the $y$-direction.

Now we discuss how to realize the quasiperiodic interaction. We consider the quasiperiodic interaction:
\begin{align}
J_{j,j+1}=J_0[1+\mu\cos(2\pi j\alpha+\delta)],\label{eq:definition_of_quasi_periodic_interaction}
\end{align}
where $J_0\equiv C_3/a_0^3$  (typical values are shown in Appendix~\ref{app:ene-emi scale}), $a_0$ is a reference length scale of the system, and $\mu$ and $\delta$ are real constants. Here $\alpha$ determines the period of the interaction, and when it is irrational, it serves as the source of disorder. In this paper we set $\alpha = (\sqrt{5} - 1)/2$. Our strategy for realizing the quasiperiodic interaction is determining the atomic positions so that $J_{ij}$ becomes a desired form. First, we set the first atom at the position $(x_1,y_1)=(0,0)$ and fix the values $\mu$ and $\delta$, where $\delta$ is randomly chosen from a uniform distribution over $[0,2\pi]$. Next we numerically solve the following equation to determine the position of the second atom by the bisection method:
\begin{align}
\frac{C_3}{R_{1,2}^3}(1-3\cos^2\theta_{1,2})=J_0\left[1+\mu\cos(2\pi \alpha+\delta)\right].\label{eq:equation_for_quasiperiodic_interaction_j=1}
\end{align}
Since Eq.~(\ref{eq:equation_for_quasiperiodic_interaction_j=1}) has two unknown variables, we restrict the range of $\theta_{1,2}$ and $R_{1,2}$ to $0\le \theta_{1,2}\le \pi/2$ and $R_{1,2}/a_0=1/(0.5l)^{1/3}$, where $l=1,2,\ldots,250$. From the solution of Eq.~(\ref{eq:equation_for_quasiperiodic_interaction_j=1}), the position of the second atom is given by $(x_2,y_2)=(x_1,y_1)+(R_{1,2}\sin\theta_{1,2}, R_{1,2}\cos\theta_{1,2})$. We iterate this procedure for $j=2,3,\,\ldots,L-1$, where $L$ is the total number of Rydberg atoms. The following equation determines the position of the $(j+1)$th atom:
\begin{align}
\frac{C_3}{R_{j,j+1}^3}(1-3\cos^2\theta_{j,j+1})=J_0\left[1+\mu\cos(2\pi \alpha j+\delta)\right].\label{eq:equation_for_quasiperiodic_interaction_j}
\end{align}
For $j\ge2$, the solution of Eq.~(\ref{eq:equation_for_quasiperiodic_interaction_j}) is not unique. To obtain a unique solution, we use the criterion of choosing the atomic position that minimizes the strength of the next-nearest-neighbor interaction $|J_{j-1,j+1}|$. 
For the initial step of determining the second atom, this criterion is unavailable. In this case, we first set $l=250$ and search for the solution of Eq.~(\ref{eq:equation_for_quasiperiodic_interaction_j=1}). If the solution exists, we accept this solution; otherwise, we set $l=249$ and search for the solution of Eq.~(\ref{eq:equation_for_quasiperiodic_interaction_j=1}). We repeat this procedure until we find a solution.
Figure~\ref{schematicpic}(c) shows an example of the atomic positions for $L=10$ and $\mu=2.5$. Although our procedure intends to make the nearest-neighbor interaction quasiperiodic, the quasiperiodicity appears even in the longer-range terms. We discuss this in Appendix \ref{app:Longer_range_terms}.

We remark here on related work on quasiperiodicity in the Rydberg-atom quantum simulator. In Ref.~\cite{Zhou2023-ve}, the authors proposed a method to realize the system with quasiperiodicity. Their method is based on modifying the interaction between the Rydberg atoms by the laser-assisted dipole-dipole interaction~\cite{Yang2022-wi}. Compared to this work, our method is based on simply tuning the position of the atoms.

Next we discuss the spin-1 case. To realize the spin-1 system, we consider arrays of $^{87}$Rb atoms arranged in the $xy$ plane. It is known that ${}^{87}$Rb atoms have F\"orster resonant channels~\cite{Ravets2014-up,Ravets2015-bz,Chew2022-ge}. The energy difference of the pair states $\ket{41F_{7/2},m_J=7/2}\otimes\ket{45P_{3/2},m_J=3/2}$ and $\ket{43D_{5/2},m_J=5/2}^{\otimes 2}$ is accidentally small. In the following, we abbreviate the states $\ket{45P_{3/2},m_J=3/2}, \ket{43D_{5/2},m_J=5/2}$, and $\ket{41F_{7/2},m_J=7/2}$ as $\ket{45P}$, $\ket{43D}$, and $\ket{41F}$, respectively. These states are coupled via the terms $\hat{d}_1^+\hat{d}_2^-+\hat{d}_1^-\hat{d}_2^+$ included in the dipole–dipole interaction. The resonance can be characterized by the F\"orster defect $\Delta E\equiv E_P+E_F-2E_D$, where $E_F$, $E_D$, and $E_P$ are the energies of the states $\ket{41F}$, $\ket{43D}$, and $\ket{45P}$, respectively. In the ${}^{87}$Rb case, the F\"orster defect becomes $\Delta E\simeq -h\times 8~{\rm MHz}$~\cite{Chew2022-ge}, which is small compared to the other energy scales. Here we introduce the shorthand notation $\ket{p}\equiv \ket{45P}$, $\ket{d}\equiv \ket{43D}$, and $\ket{f}\equiv \ket{41F}$. The dipole-dipole interaction between the two atoms separated by a distance $R$, in the subspace spanned by $\{\ket{p},\ket{d},\ket{f}\}\otimes\{\ket{p},\ket{d},\ket{f}\}$, can be approximated as:
\begin{equation}
  \begin{aligned}[b]
    \hat{V}_{\text{dd}}=\frac{C_3(\theta)}{R^{3}}&\left(\ket{dd}\bra{pf}+\ket{dd}\bra{fp}\right.
      \\&\left. +\ket{dp}\bra{pd}+\ket{df}\bra{fd} +{\rm H.c}.\right),
  \end{aligned}
  \label{spin1dip}
\end{equation}
where {\rm H.c.} denotes the Hermitian conjugate. Here, we neglected some terms, which arise from the fact that the dipole matrix elements between $\ket{p}$ and $\ket{d}$, and between $\ket{d}$ and $\ket{f}$ are not equal. As shown in Appendix~\ref{appA}, the contributions of the neglected terms are small compared to Eq.~(\ref{spin1dip}). We assign the spin-1 basis of $\ket{+}, \ket{0}$, and $\ket{-}$ as $\ket{p}, \ket{d}$, and $\ket{f}$, respectively.
The Hamiltonian of the system can be written as
\begin{equation}
  \hat{H}=\sum_{i<j}J_{ij}(\hat{\tau}^x_i\hat{\tau}^x_j+\hat{\tau}^y_i\hat{\tau}^y_j) + D\sum_i{(\hat{\tau}_i^z)^2},
  \label{Spin1Ham}
\end{equation}
where $J_{ij}$ is defined by Eq.~(\ref{eq:definition_of_Jij}), $\hat{\tau}_i^\mu\;(\mu=x,y,z)$ is the spin-1 operator at the $i$th site, and $D=\Delta E/2$ is the coefficient of the quadratic Zeeman term, which comes from the single-atom energy. The quasiperiodic interaction can be implemented similarly to the spin-1/2 case. Therefore, we can realize the $S=1/2$ and $S=1$ spin chains with quasiperiodic interactions in the Rydberg-atom quantum simulator.

\begin{figure}[H]
\centering
\includegraphics[width=\linewidth]{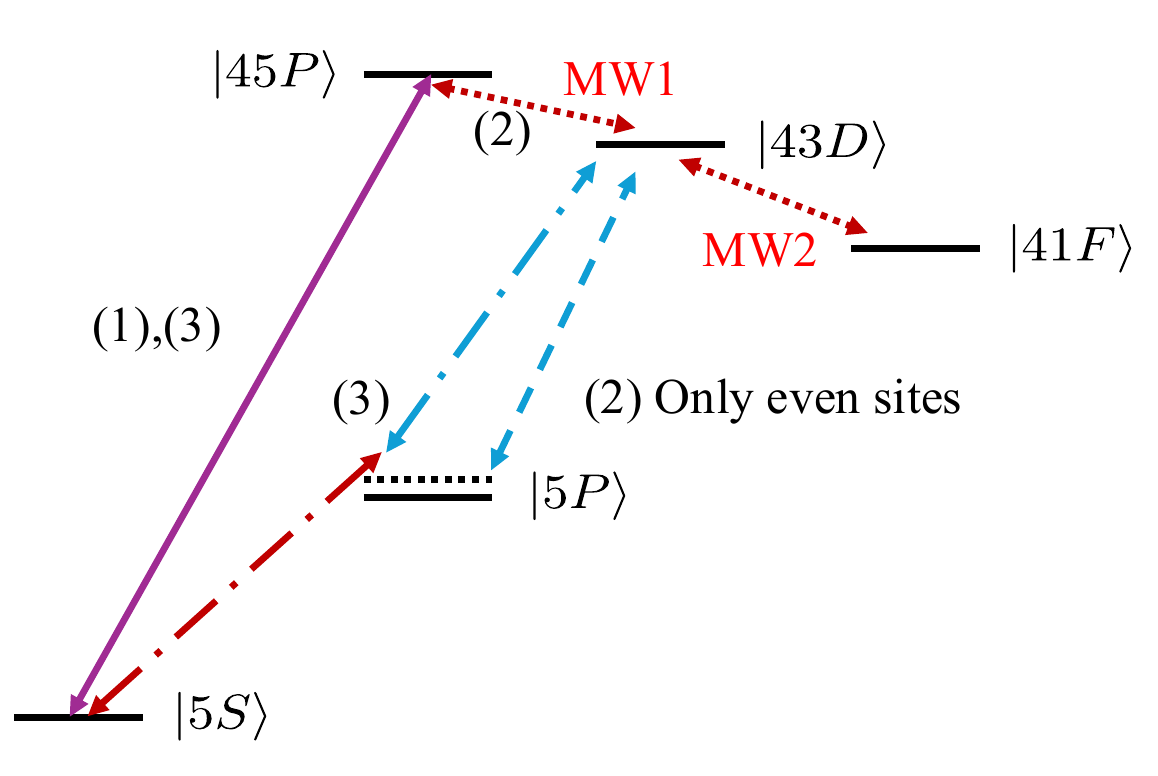}
\caption{Schematic of our procedure for initial-state preparation and measurement in the spin-1 system. The solid purple line represents UV light driving the transition $\ket{5S}\to \ket{45P}$. The red dotted lines represent the microwave transitions $\ket{45P}\to\ket{43D}$ and $\ket{43D}\to\ket{41F}$. The blue dotted lines represent the off-resonant laser applied only to the even sites. The red and blue dash-dotted lines represent the lasers used to drive the two-photon transition $\ket{43D}\to\ket{5P}\to\ket{5S}$. All parenthetical numbers correspond to the procedures explained in the text.}
\label{Statprep}
\end{figure}

Before closing this section, we discuss a possible method for preparing and observing the spin states in the spin-1 systems. As discussed in Sec.~\ref{Dyna}, the proposed protocol requires the preparation of a N\'eel-like initial spin state $\ket{-+-+\cdots}=\ket{fpfp\cdots}$, which is ideal for studying nonergodic behavior. In addition, we need to observe the staggered magnetization, defined by $M\equiv \sum_j(-1)^j\langle\hat{\tau}_j^z\rangle/L$. To obtain the staggered magnetization, the measurement of the local $z$ component of the spin is required. However, these capabilities have not yet been realized in spin-1 systems. For example, in previous experimental work by Chew {\it et al}.~\cite{Chew2022-ge}, they prepared the initial state $\ket{dd}$ and measured the population of the $\ket{d}$ state.

To overcome this problem, we propose a method to prepare and observe the spin states in spin-1 systems. The proposed sequence is schematically summarized in Fig.~\ref{Statprep}. (1) To prepare the N\'{e}el-like state, one could illuminate the atoms with an ultraviolet (UV) laser (297~{\rm nm})~\cite{Guardado-Sanchez2018-ev,Guardado-Sanchez2021-je,Hollerith2022-RD,srakaew2023subwavelength,hollerith2024Rydberg_molecule} to excite the atoms from the $\ket{5S}$ state to the $\ket{45P}$ state via a single-photon transition. According to Ref.~\cite{srakaew2023subwavelength}, coherent Rabi oscillations between $\ket{5S}$ and $|44P_{3/2}\rangle$ have been realized with a Rabi frequency of $\Omega/2\pi=1.22~\mathrm{MHz}$. (2) After preparing the $\ket{pppp\ldots}$ state, a microwave pulse would be applied to drive the transition from $\ket{45P}$ to $\ket{43D}$ ($62.528~{\rm GHz}$). Another microwave pulse would be applied subsequently to drive the transition from $\ket{43D}$ to $\ket{41F}$ ($62.536~{\rm GHz}$). During this sequence, an off-resonant laser coupling $\ket{43D}$ to $\ket{5P}$ could be applied only to the even sites to generate a staggered ac Stark shift and suppress the $\ket{45P}\to\ket{43D}$ transition. As a result, only the atoms on the odd sites would be transferred to the $\ket{41F}$ state, yielding the target state $\ket{fpfp\ldots}$. (3) We next outline a possible detection scheme for the spin states. Our scheme is based on the mid-circuit techniques~\cite{Norcia2023-Yb,Graham2023-dec,Lis2023-nov}. In the optical tweezer experiments, the population of the Rydberg states can be extracted by the recapture probability after depopulating one of the spin components through a short-lived intermediate state. Following this strategy, one could drive the $\ket{45P}\to\ket{5S}$ transition using the UV laser and drive the two-photon transition $\ket{43D}\to \ket{5P}\to \ket{5S}$ using two additional lasers. By choosing the polarization of the lasers appropriately, the atoms would finally populate two different hyperfine states in the ground-state manifold. Measuring the population of these hyperfine states would then allow one to infer the expectation value $\langle\hat{\tau}_j^z\rangle$. We note that a detection scheme without the mid-circuit technique has been demonstrated experimentally~\cite{qiao2025realization}. In that study the authors used $\ket{61S}$, $\ket{60P_{3/2}}$, and $\ket{60S}$ and measured the populations of $\ket{61S}$ and $\ket{60S}$ in separate experimental runs. Then the expectation value $\langle\hat{\tau}_j^z\rangle$ was reconstructed by these populations.

Finally, we comment on the experimental bottlenecks of the spin-1 system. First, to prepare the $\ket{pppp\ldots}$ state, we require a $\pi$-pulse of the UV laser with a sufficiently high Rabi frequency compared with the interaction between the $\ket{45P}$ states. This in turn requires a high-power UV laser, which is experimentally challenging. Second, the microwave pulses must also have sufficiently high Rabi frequencies, because the intermediate state after the first microwave pulse includes $\ket{dpdp\ldots}$. If the Rabi frequency is not high enough, the transition from $\ket{dp}$ to $\ket{pd}$ can occur due to the dipole-dipole interaction [see Eq.~(\ref{spin1dip})]. 

\section{\label{model}Model}
In the experimental setup discussed in the preceding section, we consider the $S=1/2$ and $S=1$ Hamiltonians
\begin{align}
  \hat{H}_{S=1/2} &= \sum_j \left\{J_0\left[1+\mu \cos \left(2 \pi j \alpha+\delta\right)\right] \hat{S}_j^{+} \hat{S}^{-}_{j+1}\right.\notag \\
  &\left.+ J_{j,j+2}\hat{S}_j^{+} \hat{S}^{-}_{j+2}  
  +H.c.\right\},\label{eq:S=1/2_Hamiltonian}\\
  \hat{H}_{S=1} &= \frac{1}{2}\sum_j \left\{J_0\left[1+\mu \cos \left(2 \pi j \alpha+\delta\right)\right] \hat{\tau}_j^{+} \hat{\tau}^{-}_{j+1}\right.\notag \\
  &\left.+ J_{j,j+2}\hat{\tau}_j^{+} \hat{\tau}^{-}_{j+2}  
  +H.c.\right\}+\sum^L_{j=1} D(\hat{\tau}^z_{j})^2,\label{eq:S=1_Hamiltonian}
\end{align}
where $\hat{S}^{\pm}_i\equiv \hat{S}_i^x\pm i\hat{S}_i^y$ and $\hat{\tau}^{\pm}_i\equiv \hat{\tau}^x_i\pm i\hat{\tau}_i^y$. Here, we consider the nearest- and next-nearest-neighbor interactions. The next-nearest-neighbor interaction naturally arises due to the long-range nature of the dipole-dipole interaction in our setup as shown in Sec.~\ref{setup}. For computational simplicity, we neglect longer-range interactions acting beyond the next-nearest neighbors~\cite{footnote:interaction}. The theoretical reasons for considering the next-nearest-neighbor interaction are as follows. In the case of $S=1/2$, the $XY$ model with only the nearest-neighbor interaction can be mapped to the free-fermion model via the Jordan-Wigner transformation~\cite{Jordan1928-cz}. This model is integrable, and thus it is not suitable for our purpose. In the $S=1$ case, the spin-1 $XY$ model with nearest-neighbor interaction under the open boundary conditions has a hidden $\mathrm{SU}(2)$ symmetry~\cite{Kitazawa2003-bb}. Since this symmetry is unnecessary for our purpose, we remove it by adding the next-nearest-neighbor interaction.

Our analysis uses the exact diagonalization (ED) method~\cite{Sandvik2010-va,Jung2020-es}. To perform this, we use the computational basis $\ket{\bm{\sigma}}\equiv\ket{\sigma_1\sigma_2\cdots\sigma_L}$ where $\sigma_i\in\{0,1\}$ and $\hat{S}_i^z\ket{\sigma_i}=(\sigma_i -\frac{1}{2})\ket{\sigma_i}$ in the spin-1/2 case and $\sigma_i\in\{0,1,2\}$ where $\hat{\tau}_i^z\ket{\sigma_i}=(\sigma_i -1)\ket{\sigma_i}$ in the spin-1 case.
Since the Hamiltonian has $U(1)$ symmetry, we restrict the basis $\ket{\bm{\sigma}}$ to the Hilbert subspace $\mathcal{H}\equiv\{\ket{\phi}\;| \;\hat{S}^z_{\text{tot}}\ket{\phi}=S^z_{\text{tot}}\ket{\phi}\}$ for spin-1/2 systems and $\mathcal{H}\equiv\{\ket{\phi}\;| \;\hat{\tau}^z_{\text{tot}}\ket{\phi}=\tau^z_{\text{tot}}\ket{\phi}\}$ for spin-1 systems, where we define the total magnetization as $\hat{S}^z_{\text{tot}}\equiv \sum_i\hat{S}^z_i$ for spin-1/2 systems and $\hat{\tau}^z_{\text{tot}}\equiv \sum_i\hat{\tau}^z_i$ for spin-1 systems, respectively. Throughout the rest of this paper, we use the $\mathrm{U}(1)$ symmetry to reduce the computational cost. In addition to the $\mathrm{U}(1)$ symmetry, the system has an additional $Z_2$ symmetry in the Hilbert subspace $S_{\rm tot}^z=0$ and $\tau_{\rm tot}^z=0$. We can show that the operator $\hat{\mathcal{I}}=\prod_i (2\hat{S}^x_i)$ for spin-1/2 systems and the corresponding operator $\hat{\mathcal{I}}=\prod_i\{[(\hat{\tau}^+_i)^2 + (\hat{\tau}_i^-)^2 ]/2+1-(\hat{\tau}_i^z)^2\}$ for spin-1 systems commute with the corresponding Hamiltonian in the subspace $\mathcal{H}$. We use the following Hilbert subspace in Sec.~\ref{Stat} to analyze eigenstate statistics: $\mathcal{H}_{S^z_{\text{tot}},I}=\{\ket{\phi}\;|\; \hat{S}^z_{\text{tot}}\ket{\phi}=S^z_{\text{tot}}\ket{\phi},\hat{\mathcal{I}}\ket{\phi}=I\ket{\phi}\}$ for spin-1/2 systems and $\mathcal{H}_{\tau^z_{\text{tot}},I}=\{\ket{\phi}\;|\; \hat{\tau}^z_{\text{tot}}\ket{\phi}=\tau^z_{\text{tot}}\ket{\phi},\hat{\mathcal{I}}\ket{\phi}=I\ket{\phi}\}$ for spin-1 systems.

\section{\label{results}Results}
\subsection{\label{Stat}Eigenstate statistics}
In this section we present our numerical results obtained by the ED method in the ${S}^z_{\text{tot}}=0$ and ${I}=+1$ sector for spin-1/2 systems and the $tau_{tot}^z=0$ and $I=1$ sector for spin-1 systems for each system size. In the Rydberg implementation discussed in Sec.~\ref{setup}, the coefficient of the quadratic Zeeman term is determined by the F\"{o}rster defect as $D = \Delta E/2$. In the numerical calculations, however, we treat $D$ as a model parameter and set $D=0.5J_0$.
We have checked that the nonergodic behavior discussed below is insensitive to the choice of $D$ in the parameter region considered here.

First, we show the level spacing statistics. We calculate the mean level spacing ratio or $r$-value defined as ~\cite{Atas2013-yk}
\begin{equation}
  r\equiv\left\langle\min\left( r_n,\frac{1}{r_n}\right) \right\rangle,
\end{equation}
where $r_n=s_{n+1}/{s_{n}}$ with $s_n=E_{n+1}-E_{n}$ and $E_n$ the $n$th eigenenergy ($E_{n+1}\ge E_n$) of the Hamiltonian in the zero magnetization sector, and $\langle\cdots\rangle$ represents the average over all energy eigenstates. The $r$-value is known as a useful probe of ergodicity. If the system is ergodic, the system is indistinguishable from the system described by a random matrix. In our case, since the system Hamiltonian has time-reversal symmetry, the eigenstate statistics are expected to be similar to the Gaussian orthogonal ensemble ~\cite{DAlessio2016-fh} and the $r$-value takes the value $r\simeq0.53$. In contrast, if the system is localized, the level spacing obeys the Poisson distribution and the $r$-value takes the value $r\simeq0.38$~\cite{Atas2013-yk}.

\begin{figure}[h]
\centering
\includegraphics[width=\linewidth]{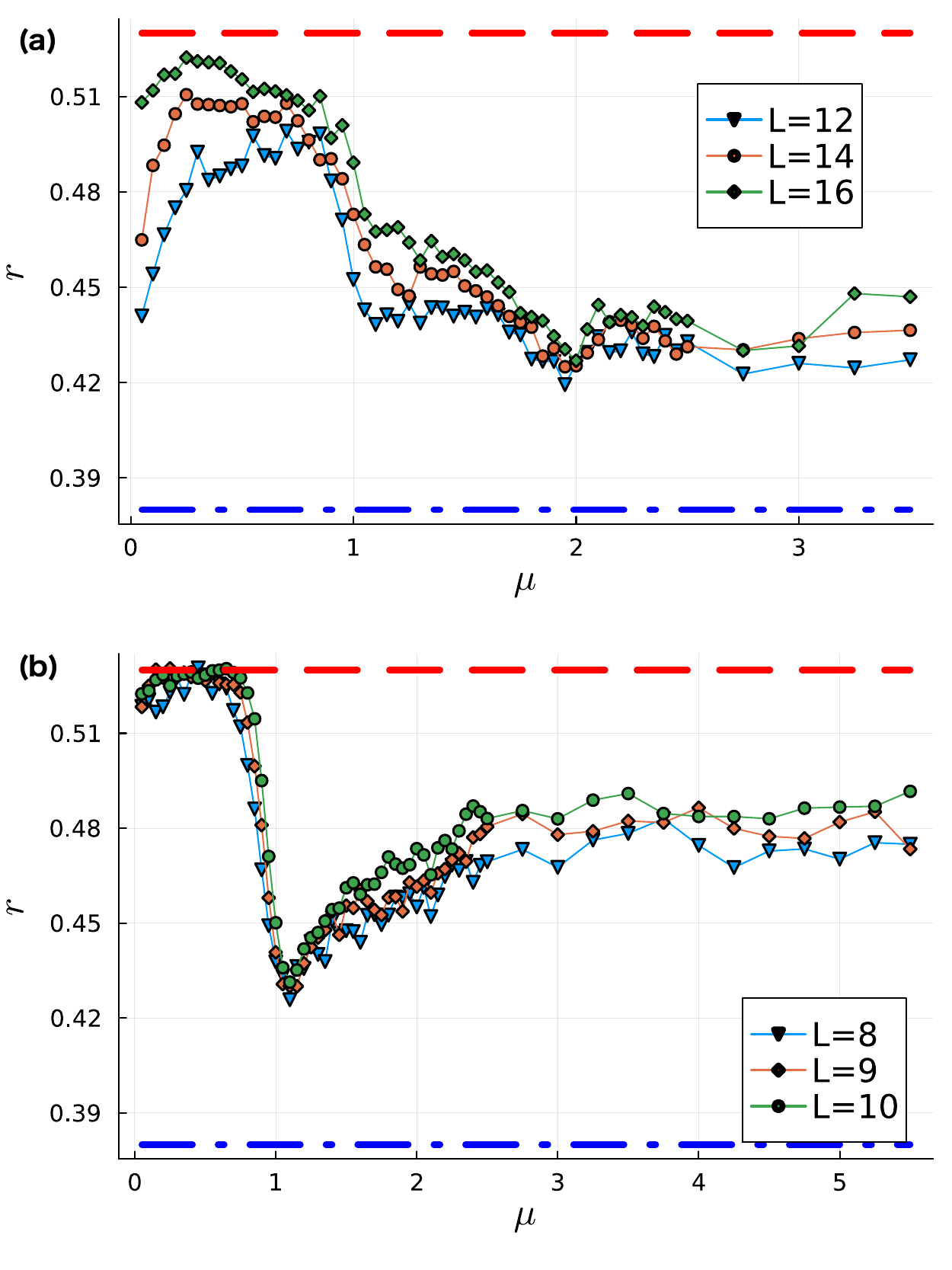}
\caption{$r$-value as a function of $\mu$. (a) $r$-value in spin-1/2 systems with system sizes $L=12,14$, and $16$ in the sector $S^z_{\text{tot}}=0$ and $I=1$. The value $\mu$ is varied up to $\mu=3.5$. (b) $r$-value in spin-1 systems in the sector $\tau_{\rm tot}^z=0$ and $I=1$. The value $\mu$ is varied up to $\mu=5.5$. We calculate $r$ by taking the average over 100 samples of $\delta\sim\mathcal{U}[0,2\pi]$. The red dashed and blue dash-dotted lines represent the value of the Wigner-Dyson distribution $r\approx0.53$ and Poisson distribution $r\approx 0.38$, respectively. The error bars are omitted because the error was smaller than the size of the marker.}
\label{LSSplot}
\vspace{0.5em}
\end{figure}

Figure~\ref{LSSplot} shows the $r$-value as a function of $\mu$ for spin-1/2 and spin-1 systems for several system sizes in the Hilbert subspace $\mathcal{H}_{S_{\rm tot}^z=0, I=1}$ and $\mathcal{H}_{\tau_{\rm tot}^z=0, I=1}$, respectively.
We can see that the $r$-value is close to the Wigner-Dyson value in the small $\mu\lesssim 1$ regime. This behavior implies that the system is ergodic. In contrast, in the regime $\mu\gtrsim 1.0$, the $r$-value deviates from the Wigner-Dyson value and finally takes the value around $r\simeq0.44$ for the spin-1/2 systems, and $r\simeq0.48$ for the spin-1 systems. These values are different from the values of the Wigner-Dyson distribution and the Poisson distribution. This behavior implies that the system is neither ergodic nor completely localized. This behavior is consistent with previous studies~\cite{Wang2021-zy,Zhou2023-ve,Roy2023-DN} and thus can be regarded as evidence of an intermediate regime called many-body critical (MBC)~\cite{Wang2021-zy} or nonergodic extended regime~\cite{Tomasi2019,Roy2023-DN}.

Next, we calculate the eigenstate statistics of the von Neumann entanglement entropy (EE) defined as $S_n\equiv-\text{Tr}(\hat{\rho}_{A}\ln{\hat{\rho}_{A}})$ where the index $n$ indicates that this is the EE of the $n$th eigenstate of the Hamiltonian $\ket{E_n}$ and $\hat{\rho}_A$ is the reduced density matrix defined by tracing out the subsystem $B$: $\hat{\rho}_A\equiv{\rm Tr}_B(\ket{E_n}\bra{E_n})$. Throughout this paper, we take subsystems $A$ and $B$ to be the left and right halves of the total system. When the system is ergodic, since it is indistinguishable from random matrices, the half-chain EE obeys the volume-law scaling, in which the EE scales linearly with the system size, although ground or low-lying states do not obey this scaling. Therefore, it is natural to calculate von Neumann EE only for the eigenstates in the middle of the spectrum of the Hamiltonian. In this paper, we calculate the EE over the middle 100 eigenstates and take their average. In addition, we also average over samples with different values of $\delta$ to eliminate effects specific to a particular value of $\delta$.

\begin{figure}[t]
  \centering
  \includegraphics[width=\linewidth]{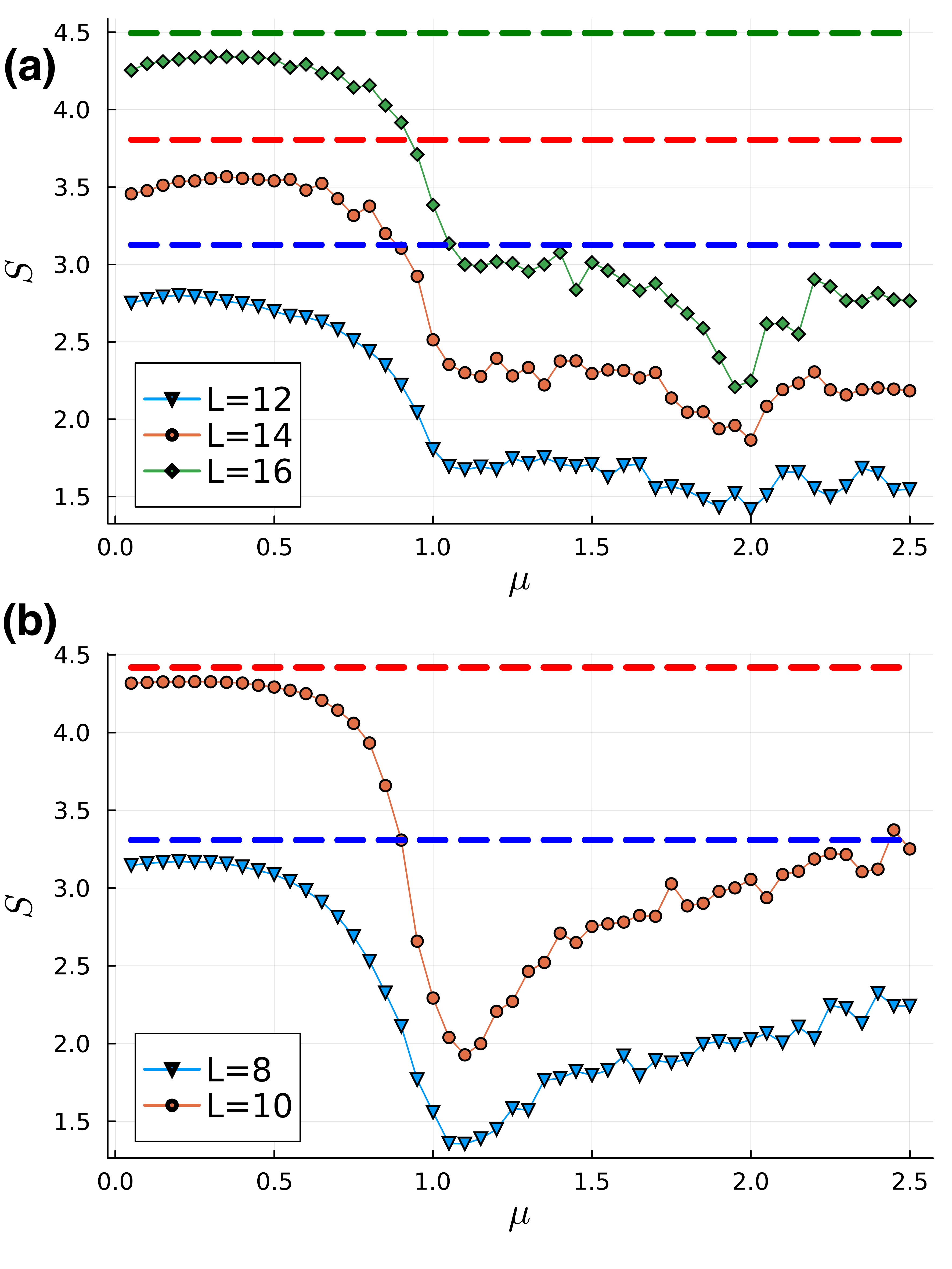}\\
  \caption{Eigenstate von Neumann entanglement entropy as a function of $\mu$. (a) Spin-1/2 systems with system sizes $L=12,14$, and $16$ in the $S^z_{\text{tot}}=0$ and $I=1$ sector. (b) Spin-1 systems with system sizes $L=8$ and $10$ in the $\tau^z_{\text{tot}}=0$ and $I=1$ sector. We calculate $S$ by taking the average over 100 samples of $\delta\sim\mathcal{U}[0,2\pi]$ and over 100 states in the middle of the spectrum. Error bars are omitted because they are smaller than the size of the marker. The blue, red, and green dotted lines represent the Page values for each system size.}
  \label{Entplot}
\end{figure}

Figure \ref{Entplot} shows the averaged EE as a function of $\mu$ for the spin-1/2 and spin-1 systems with several system sizes. We find that in the regime $\mu\ge1.0$, the EE follows the volume-law scaling. We also find that the value of the EE in the small $\mu$ regime is close to the Page value~\cite{Page1993-AE}, which is the averaged EE of the random vectors whose complex elements are independently drawn from a normal distribution. If a system is ergodic and its energy is close to the middle of the spectrum, the bipartite entanglement entropy approaches the Page value. In the large-$\mu$ regime, the value of the EE deviates from the Page value. However, the EE still obeys the volume-law scaling. This behavior differs from that of the conventional MBL case because the EE in the MBL regime obeys the area-law scaling. Therefore, our results imply that the wave function in this regime is extended but nonergodic. All of the results we have shown are consistent with previous studies on critical regimes in quasiperiodic systems~\cite{Wang2021-zy,Zhou2023-ve}. Therefore, we conclude that our proposed models exhibit unconventional localization phenomena, which are worth investigating using quantum simulators at system sizes inaccessible to  classical computers.

Finally, we show the results of the inverse participation ratio (IPR) calculated with the highly excited eigenstates. The IPR of an eigenstate $\ket{E_k}$ is defined as
\begin{align}\label{IPR4eigst}
    \operatorname{IPR}_2(E_k)=\sum_{\bm{\sigma}}|\langle {\bm{\sigma}|E_k}\rangle|^4,
\end{align}
where $\bm{\sigma}$ represents a particular spin configuration and the sum is taken over all possible spin configurations $\bm{\sigma}$. The IPR quantifies the degree of localization. When the wave function is spread over the entire Hilbert space, the IPR takes its minimum value as $\operatorname{IPR}_2 \sim 1/N$, where $N$ is the dimension of the Hilbert space.
On the other hand, if the wave function is localized around a particular basis state, the IPR behaves as $\operatorname{IPR}_2\sim 1$. In the following, we calculate the negative logarithm of the IPR, $-\ln\operatorname{IPR_2}$, rather than the IPR itself. We also average over 100 realizations of the random variable $\delta$ and about 100 eigenstates around the center of the spectrum.

\begin{figure}[h]
\centering
\includegraphics[width=\linewidth]{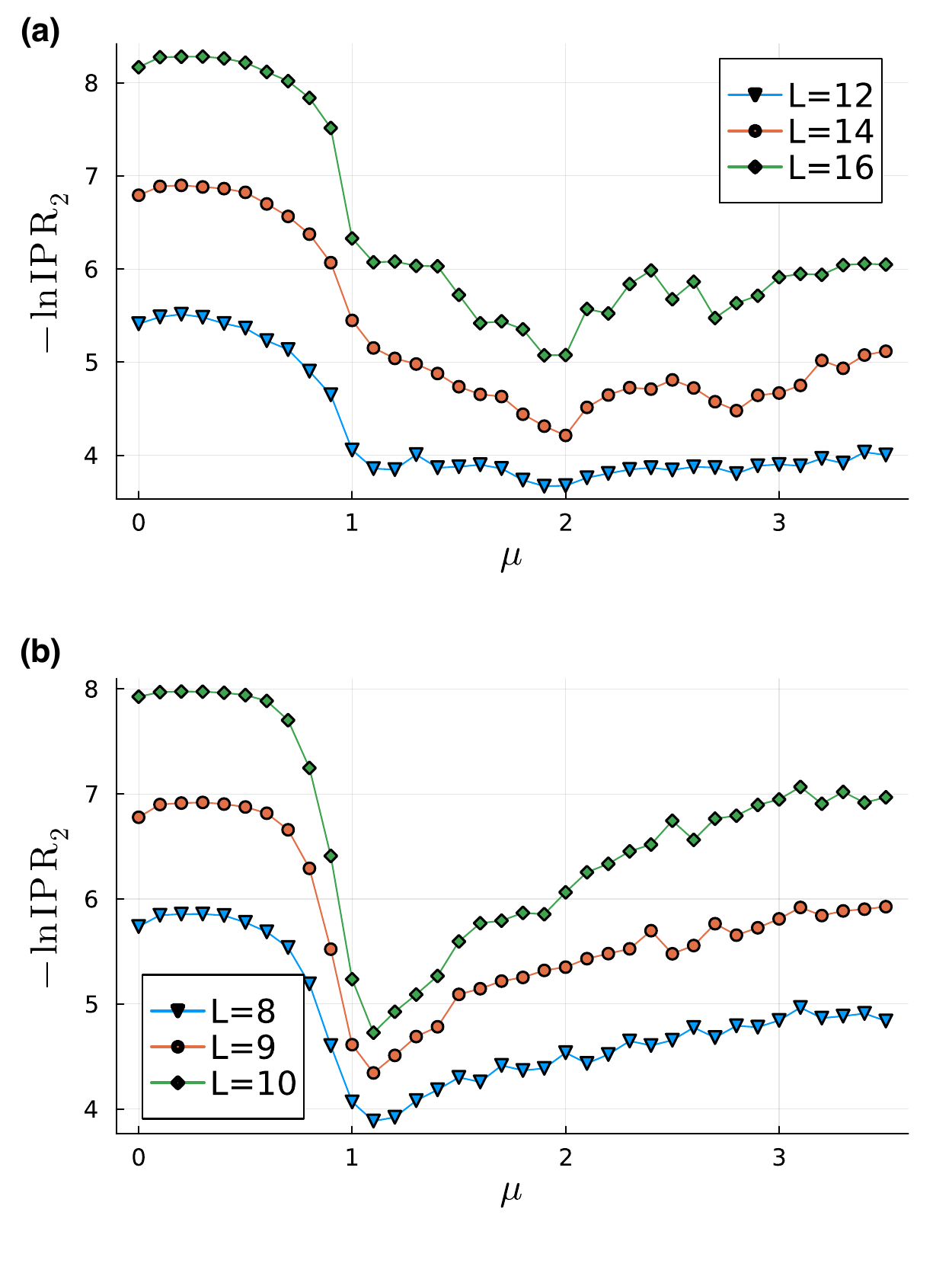}
\caption{Negative logarithm of the IPR as a function of $\mu$. (a) Spin-1/2 systems with system sizes $L=12,14$, and $16$ in the $S^z_{\text{tot}}=0$ and $I=1$ sector. (b) Spin-1 systems with system sizes $L=8,9$, and $10$ in the $\tau^z_{\text{tot}}=0$ and $I=1$ sector. We average over 100 samples of $\delta\sim\mathcal{U}[0,2\pi]$ and over 100 states in the middle of the spectrum.}\label{fig:IPRplot}
\end{figure}

Figure~\ref{fig:IPRplot} shows the averaged negative logarithm of the IPR as a function of the quasiperiodicity $\mu$ for the spin-1/2 and spin-1 systems with several system sizes. We find that, in the regime $\mu \ge 1.0$, the observable deviates from the ergodic value $\ln N$. This result implies that the eigenstate wave functions are not fully delocalized and occupy only part of the many-body Hilbert space. 

We also discuss the mobility edge of the system and the system-size dependence of the IPR and EE. The results are shown in Appendixes D and E~\ref{appB} and \ref{appC}.

\subsection{\label{Dyna}Dynamics}

In this section we show the results of real-time evolution obtained by solving the Schr\"odinger equation
\begin{equation}
  i\hbar\frac{d}{d t}\ket{\psi (t)}=\hat{H}\ket{\psi(t)},
\end{equation}
where $\ket{\psi(t)}$ is a state vector at time $t$. Throughout the paper, we set the initial state to the N\'{e}el state $\ket{\psi(0)}=\ket{\downarrow\uparrow\downarrow\uparrow\cdots}$ for the spin-1/2 system and the N\'{e}el-like state $\ket{\psi(0)}=\ket{-+-+\cdots}$ for the spin-1 systems. We use the ED method for the time evolution. We focus on two observables. The first is the staggered magnetization $M\equiv2\sum_{i=1}^L(-1)^i\langle\hat{S}_i^z\rangle/L$ for the spin-1/2 systems and $M\equiv \sum_{i=1}^L(-1)^i\langle\hat{\tau}_i^z\rangle/L$ for the spin-1 system. The staggered magnetization becomes zero in the thermal-equilibrium states; therefore, we can use it as a probe for thermalization. In addition, this quantity is relatively easy to measure in Rydberg-atom experiments. The second observable is the half-chain EE. When we start the calculation from a product state, we can see that the EE grows to the Page value if the system is ergodic. In contrast, in a localized system the EE does not reach a thermal value even after a long time, but grows logarithmically: $S\propto\ln({t})$. Note that we do not consider the next-nearest-neighbor interaction terms in the spin-1 model for computational simplicity. Since the next-nearest-neighbor interaction is small compared with the nearest-neighbor interaction, we can expect that this omission does not change the qualitative behavior (see Ref. \cite{footnote:interaction}).

\begin{figure}[t]
    \centering
\includegraphics[width=\linewidth]{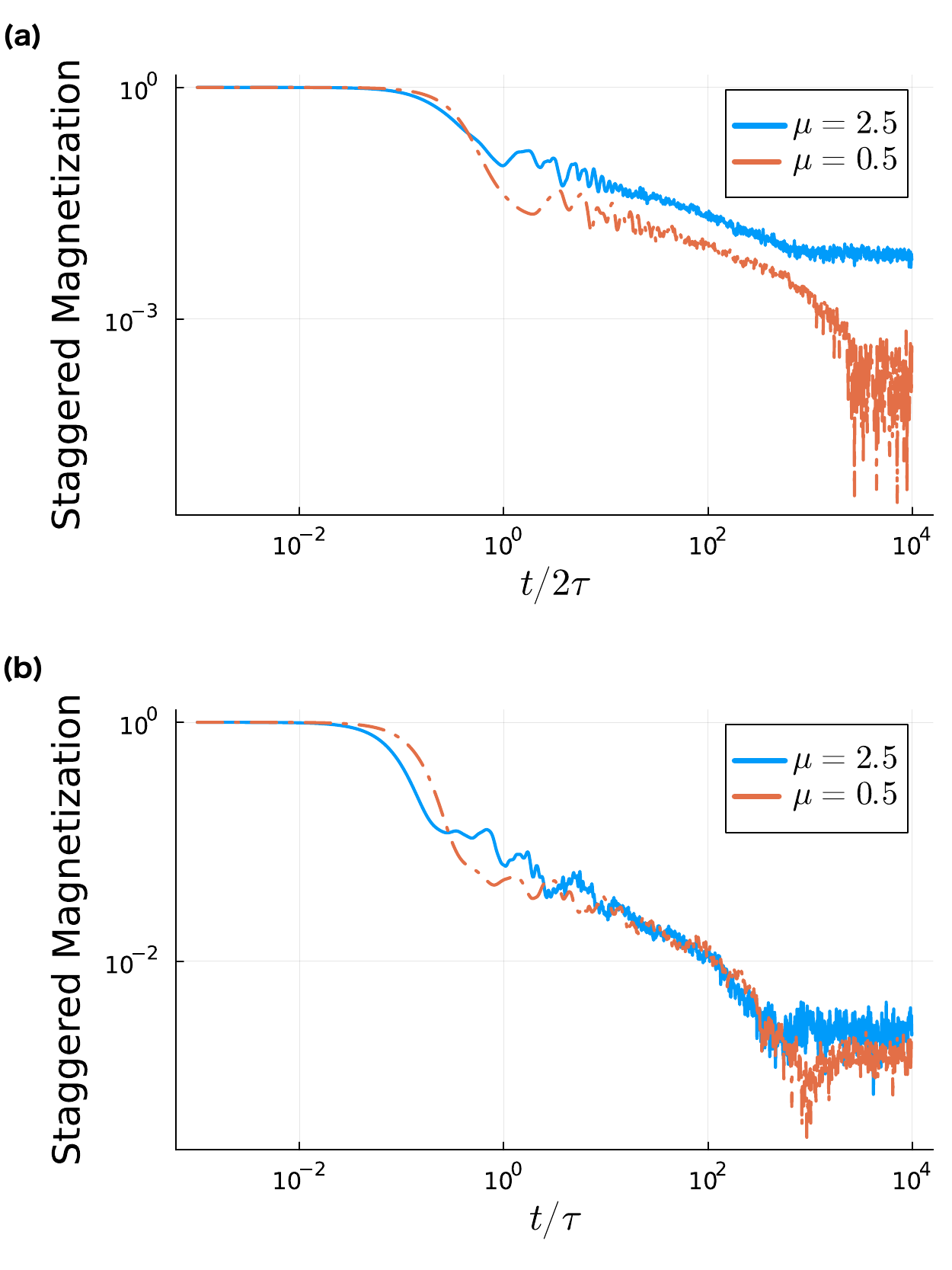}
    \caption{Time evolution of the staggered magnetization. (a) Spin-1/2 systems for $L=16$. (b) Spin-1 systems for $L=10$. Here we define $\tau\equiv \hbar/J_0$.
    }
    \label{STmag}
\end{figure}

Figure \ref{STmag} shows the time evolution of the staggered magnetization. To calculate this quantity, we average over 100 samples of $\delta\sim\mathcal{U}[0,2\pi]$. We can see that the staggered magnetization decays to zero for small $\mu$ in both spin-1/2 and spin-1 systems. In the large-$\mu$ case, the staggered magnetization decays, but it seems to remain nonzero even after a sufficiently long time. This slow decay suggests that the quasiperiodicity enhances the nonergodicity.

 \begin{figure}[h]
    \centering
\includegraphics[width=\linewidth]{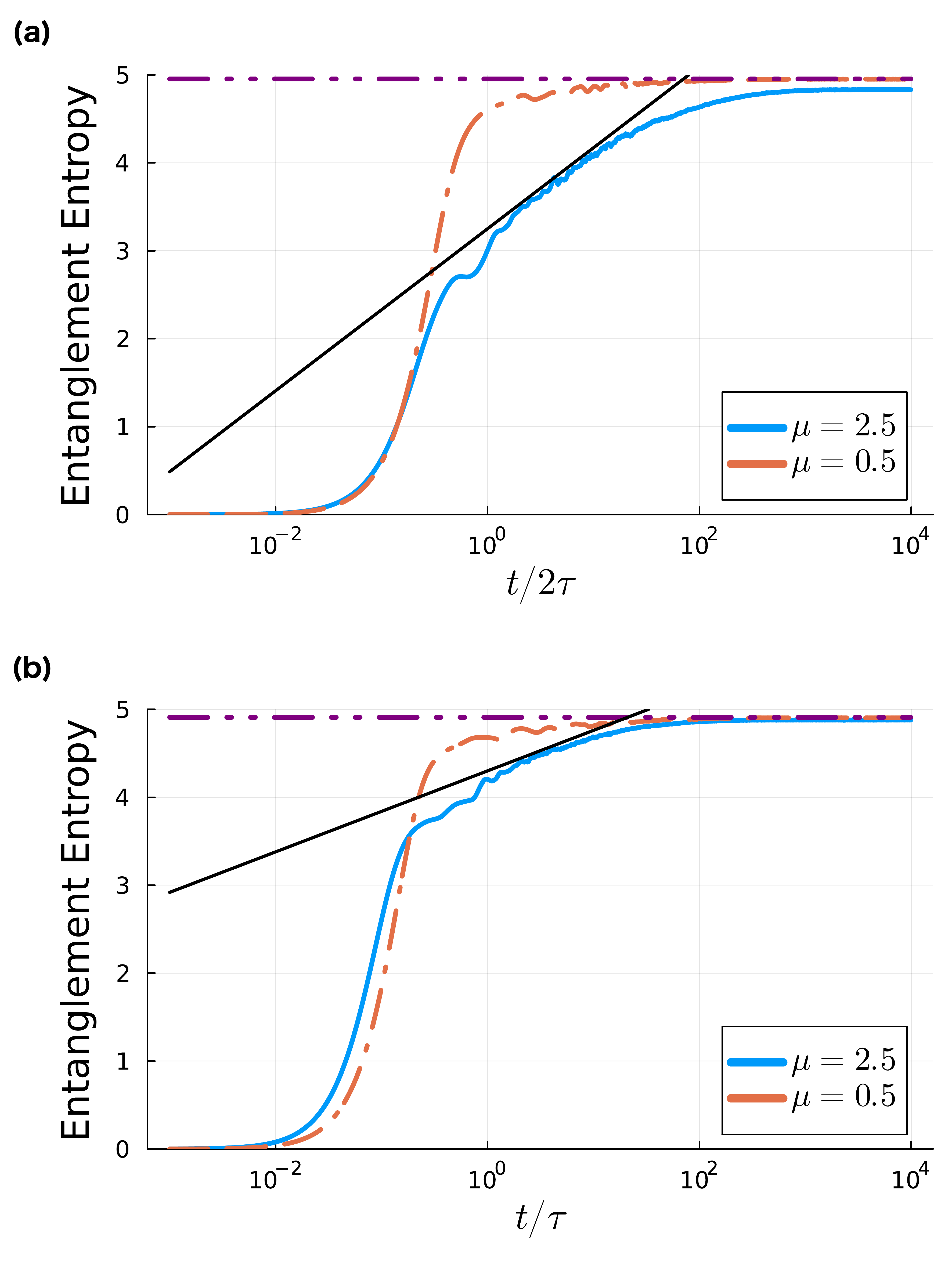}
    \caption{Time evolution of the half-chain EE. (a) Spin-1/2 systems for $L=16$. (b) Spin-1 systems for $L=10$. The black solid line represents $\ln(t)$ as a guide to the eye. The purple dash-double-dotted line represents the Page value.
    }
    \label{EEDyna}
\end{figure}

Figure \ref{EEDyna} shows the results of the half-chain EE. We can see that the EE grows rapidly and saturates at the Page value for both the spin-1/2 and spin-1 cases. In the large-$\mu$ regime, we can see logarithmic growth of the EE in the intermediate timescale. This behavior is similar to that in MBL systems. Therefore, combining the results of the eigenstate statistics, we conclude that our proposed models exhibit a many-body critical regime.

\section{\label{sum}Conclusion}
We have proposed an experimental procedure to realize the $S=1/2$ and $1$ quantum spin chains with quasiperiodic interactions using the Rydberg-atom platform. To achieve this, we exploited the anisotropic nature of the dipole-dipole interactions between the Rydberg atoms and the tunability of the atomic positions by optical tweezer techniques. We also investigated the nonergodic properties of our proposed models. From the calculations of the $r$-value, bipartite entanglement entropy, and dynamics of the staggered magnetization and the half-chain EE, we have found that these models exhibit properties different from those of the well-known ergodic and MBL regimes and are consistent with an MBC regime. However, our calculations are limited to small system sizes due to the exponentially large Hilbert space. Experiments using Rydberg-atom quantum simulators will reveal the properties of the MBC regime.

Finally, we discuss prospects for long-time experiments. To explore nonergodicity in many-body systems, one needs to perform long-time experiments. In typical Rydberg-atom experiments, the timescale is limited by the lifetime of the Rydberg states, which is roughly $10~{\rm \mu s}$. To overcome this limitation, circular Rydberg atoms are promising candidates, with lifetimes much longer than those of low–angular-momentum Rydberg states. In fact, millisecond-scale lifetimes of circular Rydberg states have been observed~\cite{cantat-Moltrecht2020long,Holzl2024long}. Recently, anisotropic dipole–dipole interactions between circular Rydberg atoms have also been observed~\cite{mehaignerie2025interacting}. Using circular Rydberg atoms, quantum simulation of nonergodic systems may become feasible. However, experimental realization of the spin-1 model remains challenging. In particular, the F\"orster resonance between circular Rydberg states required for realizing the spin-1 $XY$ interaction has not yet been observed experimentally. Nevertheless, recent theoretical studies suggest that the interaction strengths of circular Rydberg states can be tuned by external electric and magnetic fields~\cite{dobrzyniecki2025tunable}. In addition, blackbody-radiation-induced decoherence can be suppressed in cryogenic environments~\cite{cantat-Moltrecht2020long,zhang2025high}. These developments indicate that circular Rydberg atoms are a promising platform for long-time studies of the nonergodic dynamics discussed in this work.

\section{Acknowledgements}
We thank T. Tomita for useful discussions, especially regarding Rydberg-atom experiments. This work was supported by JSPS KAKENHI Grants No.~JP22H05268 and JP25K00215 (M.K.), JST ASPIRE No.~JPMJAP24C2 (M.K.), and JST SPRING Grant No. ~JPMJSP2151 (T.Y.).



\appendix
\begin{widetext}
\section{Experimental considerations regarding energy and time scales}\label{app:ene-emi scale}

In this Appendix we summarize representative physical energies and decay scales. These values are not used as additional fitting parameters in the numerical calculations; throughout the main text, energies are measured in units of \(J_0\). All the results in this Appendix were obtained using the Alkali Rydberg Calculator (ARC)~ Python library~\cite{SIBALIC2017319}.

The interaction scale is defined as
\[
J_0 = \frac{C_3}{a_0^3},
\qquad
\frac{J_0}{h} = \frac{C_3/h}{a_0^3},
\qquad
\tau \equiv \frac{\hbar}{J_0}.
\]

The spontaneous-emission rate of a Rydberg state \(|r\rangle\) is defined as
\[
\Gamma_{\rm sp}^{(r)} = \sum_{f} A_{r\to f},
\]
where \(A_{r\to f}\) is the Einstein \(A\) coefficient for spontaneous decay
from \(|r\rangle\) to a lower-lying state \(|f\rangle\).

Below, we present the results for the two cases considered in the main text: $S=1/2$ and $S=1$. 

For the $S=1/2$ case, assuming that the $\ket{60S_{1/2},m_J=1/2}$ and $\ket{60P_{1/2},m_J=-1/2}$ states of $^{87}$Rb are chosen as the Rydberg states, as in Ref.~\cite{De_Leseleuc2019-xu}, we obtain
\begin{align}
    &\frac{|C_3|}{h}=1.518 \,\,{\rm GHz\cdot\mu m^3},\quad \frac{|J_0|}{h}=56.21 \,\,{\rm MHz} \,\,(\text{with typical distance $a_0= 3\,\,{\mu m}$}),\nonumber\\
    & \frac{\Gamma_{\rm sp}^{(S)}}{2\pi} = 0.692~{\rm kHz},\quad \frac{\Gamma_{\rm sp}^{(P)}}{2\pi} = 0.319~{\rm kHz}.
\end{align}

For the $S=1$ case, as proposed in the main text, we choose three Rydberg states $\ket{45P_{3/2},m_J=3/2}, \ket{43D_{5/2},m_J=5/2}$, and $\ket{41F_{7/2},m_J=7/2}$. We then obtain
\begin{align}
    &\frac{|C_3|}{h}=0.374 \,\,{\rm GHz\cdot\mu m^3},\quad \frac{|J_0|}{h}=13.84 \,\,{\rm MHz} \,\,(\text{with typical distance $a_0= 3\,\,{\mu m}$}),\nonumber\\
    & \frac{\Gamma_{\rm sp}^{(45P)}}{2\pi} = 0.850~{\rm kHz},\quad \frac{\Gamma_{\rm sp}^{(43D)}}{2\pi} = 2.009~{\rm kHz},\quad \frac{\Gamma_{\rm sp}^{(41F)}}{2\pi} = 3.370~{\rm kHz}.
\end{align}

\section{Quasiperiodic signatures in longer-range interactions}\label{app:Longer_range_terms}
In this Appendix, we discuss the quasiperiodicity of the longer-range interaction terms. As shown in Sec.~\ref{setup}, our scheme is designed to make the nearest-neighbor interaction quasiperiodic. However, the longer-range interaction terms are not necessarily quasiperiodic.

To clarify the quasiperiodicity of the longer-range interaction terms, we investigate the Fourier spectrum of the interaction
\begin{align}
g(\kappa)\equiv\sum_{j=1}^{L-1}e^{2\pi i \kappa j/(L-1)}J_{j,j+r}/J_0,
\label{eq:definition_of_Fourier_spectrum_of_interactions}
\end{align}
where $-(L-1)/2\le \kappa< (L-1)/2$ for odd $L$, $-(L-2)/2\le \kappa\le (L-2)/2$ for even $L$, and $r=1,2,\ldots$ denotes the interaction range. For $j+r>L$, we set $J_{j,j+r}=0$.

We show the numerical results for the Fourier spectrum $|g(\kappa)|$ for various values of $\mu$ in Fig.~\ref{fig:fourier_spec}. Since the nearest-neighbor interaction is of the form $1+\mu\cos(2\pi\alpha j+\delta)$, peaks in the Fourier spectrum appear around the corresponding wave numbers. Interestingly, we find that the same peaks also appear in the longer-range interactions. These results imply that our scheme captures the quasiperiodicity of the longer-range interactions, despite not being explicitly designed to make them quasiperiodic.

\begin{figure}
    \centering
    \includegraphics[width=0.75\linewidth]{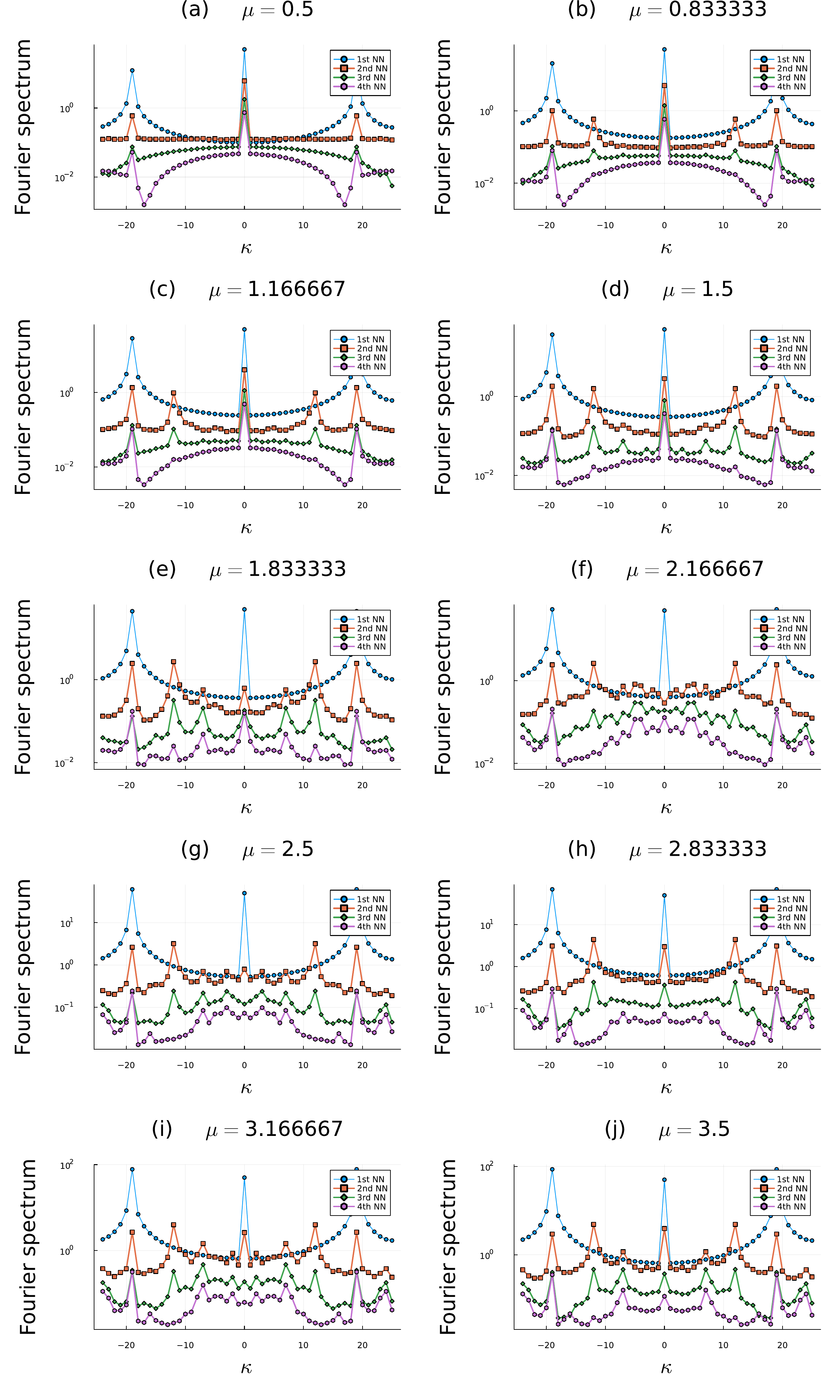}
    \caption{Fourier spectrum $|g(\kappa)|$ of $r$-th-nearest-neighbor interactions with $r=1,2,3,4$. All results here are calculated with $L=50$ by averaging over $100$ samples for (a) $\mu=0.5$, (b) $\mu=0.833333$, (c) $\mu=1.166667$, (d) $\mu=1.5$, (e) $\mu=1.833333$, (f) $\mu=2.166667$, (g) $\mu=2.5$, (h) $\mu=2.833333$, (i) $\mu=3.166667$, and (j) $\mu=3.5$.}
    \label{fig:fourier_spec}
\end{figure}

\section{\label{appA} Details of the spin-1 $XY$ Hamiltonian}

In this appendix we discuss the derivation of Eq.~(\ref{Spin1Ham}) in detail. As discussed around Eq.~(\ref{spin1dip}), the $^{87}$Rb atom possesses three almost equispaced Rydberg levels. We start by considering two Rydberg atoms interacting with each other. The interaction Hamiltonian is given by
\begin{equation}
\begin{aligned}
\hat{V}_{{\rm dd}} & =\frac{d_{p d} d_{d p}}{4 \pi \varepsilon_0R^3}\frac{1-3 \cos ^2 \theta}{2}(|p d\rangle\langle d p|+| d p\rangle\langle p d|) \\
& +\frac{d_{f d} d_{d f}}{4 \pi \varepsilon_0R^3}\frac{1-3 \cos ^2 \theta}{2}(|d f\rangle\langle f d|+| f d\rangle\langle d f|) \\
& +\frac{d_{f d} d_{p d}}{4\pi\varepsilon_0R^3}\frac{1-3 \cos ^2 \theta}{2}(|d d\rangle\langle f p|+|d d\rangle\langle p f|+| p f\rangle\langle d d|+| f p\rangle\langle d d|).
\end{aligned}
\end{equation}
Here, we define dipole matrix elements as
\begin{equation}
    d_{pd}\equiv\bra{p}\hat{d}_-\ket{d},\quad d_{dp}\equiv\bra{d}\hat{d}_+\ket{p},\quad d_{fd}\equiv\bra{f}\hat{d}_+\ket{d},\quad d_{df}\equiv\bra{d}\hat{d}_-\ket{f},
\end{equation}
where $\hat{d}_{+}(\hat{d}_{-})$ represents the $+(-)$ component of the dipole operator.

Now we move to the pseudo-spin representation. From the mapping $\ket{p}\rightarrow\ket{+},\ket{d}\rightarrow\ket{0}$, and $\ket{f}\rightarrow\ket{-}$, the interaction Hamiltonian becomes
\begin{equation}
\begin{split}
    \hat{V}_{{\rm dd}}=&J_1[\ket{00}(\bra{+-}+\bra{-+})+(\ket{+-}+\ket{-+})\bra{00}]\\
    &\;+J_2(\ket{0-}\bra{-0}+\ket{-0}\bra{0-})+J_3(\ket{0+}\bra{+0}+\ket{+0}\bra{0+}),
\end{split}
\end{equation}
where $J_1\equiv{d_{f d} d_{p d}}(1-3\cos^2\theta)/{8\pi\varepsilon_0R^3},J_2\equiv{d_{f d} d_{d f}}(1-3\cos^2\theta)/{8 \pi \varepsilon_0R^3}$, and $J_3\equiv{d_{p d} d_{d p}}(1-3\cos^2\theta)/{8 \pi \varepsilon_0R^3}$. To obtain the operator form of the interaction Hamiltonian, we use the relations
\begin{equation}
\begin{split}
    &|0\rangle\langle 0|=\hat{1}-\left(\hat{\tau}^z\right)^2, \quad|+\rangle\langle-|=(\hat{\tau}^{+})^2 / 2, \quad|-\rangle\langle+|=\left(\hat{\tau}^{-}\right)^2 / 2, \\
&|+\rangle\langle+|=\frac{1}{2}\left[\left(\hat{\tau}^z\right)^2+\hat{\tau}^z\right], \quad|-\rangle\langle-|=\frac{1}{2}\left[\left(\hat{\tau}^z\right)^2-\hat{\tau}^z\right], \\
&|+\rangle\langle 0|=\frac{1}{2 \sqrt{2}}\left(\hat{\tau}^{+}+\left\{\hat{\tau}^{+}, \hat{\tau}^z\right\}\right), \quad|0\rangle\langle+|=\frac{1}{2 \sqrt{2}}\left(\hat{\tau}^{-}+\left\{\hat{\tau}^{-}, \hat{\tau}^z\right\}\right), \\
&|0\rangle\langle-|=\frac{1}{2 \sqrt{2}}\left(\hat{\tau}^{+}-\left\{\hat{\tau}^{+}, \hat{\tau}^z\right\}\right), \quad|-\rangle\langle 0|=\frac{1}{2 \sqrt{2}}\left(\hat{\tau}^{-}-\left\{\hat{\tau}^{-}, \hat{\tau}^z\right\}\right),\end{split}
\end{equation}
where $\{\hat{A}, \hat{B}\}\equiv \hat{A}\hat{B}+\hat{B}\hat{A}$ is the anticommutator. From these relations, we obtain
\begin{equation}
    \begin{aligned}
        \hat{V}_{{\rm dd}}&=\left(\frac{J_1}{4}+\frac{J_2+J_3}{8}\right)\left(\hat{\tau}_1^+\hat{\tau}_2^-+\hat{\tau}_1^-\hat{\tau}_2^+\right)\\
        &+\left(-\frac{J_1}{4}+\frac{J_2+J_3}{8}\right)\left(\left\{\hat{\tau}_1^{+}, \hat{\tau}_1^z\right\}\left\{\hat{\tau}_2^{-}, \hat{\tau}_2^z\right\}+\left\{\hat{\tau}_1^{-}, \hat{\tau}_1^z\right\}\left\{\hat{\tau}_2^{+}, \hat{\tau}_2^z\right\}\right)\\
        &-\frac{J_2-J_3}{8}\left(\hat{\tau}_1^{+}\left\{\hat{\tau}_2^{-}, \hat{\tau}_2^z\right\}+\hat{\tau}_1^{-}\left\{\hat{\tau}_2^{+}, \hat{\tau}_2^z\right\}+\left\{\hat{\tau}_1^{-}, \hat{\tau}_1^z\right\} \hat{\tau}_2^{+}+\left\{\hat{\tau}_1^{+}, \hat{\tau}_1^z\right\} \hat{\tau}_2^{-}\right).
    \end{aligned}
    \label{IntHamA4}
\end{equation}
When $J_1=J_2=J_3\equiv J$, the interaction becomes the $XY$ model,
\begin{equation}
    \hat{V}_{{\rm dd}}=\frac{J}{2}(\hat{\tau}_1^+\hat{\tau}_2^- + \hat{\tau}_1^-\hat{\tau}_2^+)=J(\hat{\tau}_1^x\hat{\tau}_2^x + \hat{\tau}_1^y\hat{\tau}_2^y).
\end{equation}
We note that the deviation from the pure $XY$ model is relatively small in the actual experimental situation~\cite{Chew2022-ge}. In fact, we evaluate the ratios of the strengths of the other terms to that of the $XY$ term. In the case of Ref.~\cite{Chew2022-ge}, we obtain $|J_{{\rm sub}1}/J_{XY}|\sim 0.0007$ and $|J_{{\rm sub}2}/J_{XY}|\sim 0.03$, where $J_{XY}\equiv J_1/4+(J_2+J_3)/8$, $J_{{\rm sub}1}\equiv -J_1/4+(J_2+J_3)/8$, and $J_{{\rm sub}2}\equiv -(J_2-J_3)/8$. Therefore, we neglect this deviation in the main text. Here, we add the single-atom terms and consider an $L$-atom chain. The total Hamiltonian then becomes
\begin{equation}
      H=\sum_{i<j}J_{ij}(\hat{\tau}^x_i\hat{\tau}^x_j+\hat{\tau}^y_i\hat{\tau}^y_j) + D\sum_i{(\hat{\tau}_i^z)^2} + p\sum_i \hat{\tau}_i^z,\label{eq:Hamiltonian_S=1_appendix}
\end{equation}
where $p\equiv (E_P-E_F)/2$ is the linear Zeeman energy and $D\equiv (E_P+E_F-2E_D)/2$ is the quadratic Zeeman energy. Although Eq.~(\ref{eq:Hamiltonian_S=1_appendix}) contains the linear Zeeman term, this can be removed by a unitary transformation. Owing to the $U(1)$ symmetry of the system, this term does not affect the physics.

\section{Mobility edge analysis using the $r$ value}\label{appB}

In this appendix we perform a detailed analysis to detect the mobility edge of our systems.

To explore the mobility edge of our system, we show the results of calculations in which the $r$ value is resolved across the spectrum. In Fig.~\ref{figS1:Mobility_edge} we plot the $r$ value as function of $\mu$ and normalized eigenenergy $\varepsilon$, where $\varepsilon\equiv (E-E_0)/(E_{\rm max}-E_0)$, and $E_{0}$ and $E_{\rm max}$ are the ground-state and maximum energies, respectively. Here, we divide the eigenenergies of the Hamiltonian into $20$ equal parts and calculate the $r$ value for each interval for fixed $\mu$. Within the system sizes studied here, we find no signature of a mobility edge in either the $S=1/2$ or $S=1$ model. These results are consistent with the previous study~\cite{Wang2021-zy}, which investigated different models thought to exhibit similar phenomena, making this interpretation reasonable.

\begin{figure}[H]
    \centering
    \includegraphics[width=\linewidth]{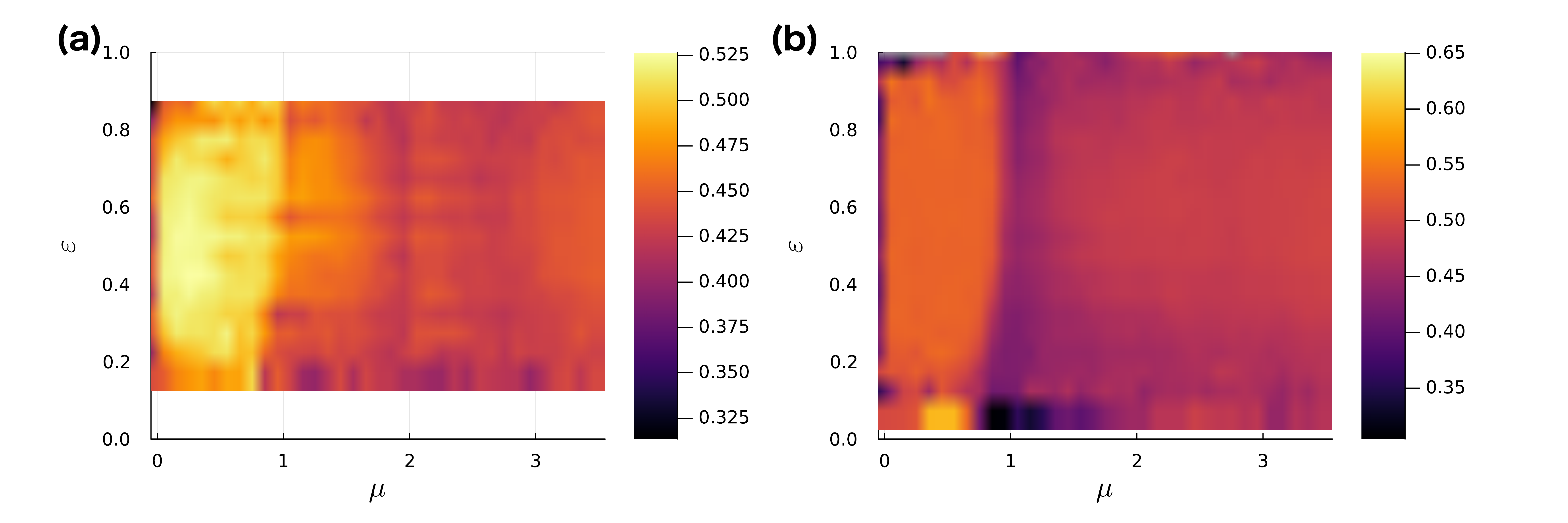}
    \caption{ $r$ value as a function of $\mu$ and $\varepsilon$. (a) $r$ value in spin-1/2 systems with system size $L=16$ in the sector $S^z_{\text{tot}}=0$ and $I=1$. (b) $r$ value in spin-1 systems in the sector $\tau_{\rm tot}^z=0$ and $I=1$ with system size $L=11$. We calculate $r$ by taking the average over 100 samples of $\delta\sim\mathcal{U}[0,2\pi]$.
    }
    \label{figS1:Mobility_edge}
\end{figure}

\section{System-size dependence of the inverse participation ratio and entanglement entropy}\label{appC}

In this appendix, in order to infer the properties of the Hamiltonian in the thermodynamic limit from our finite-size calculation results, we present the scaling of the observables computed in the main text as the system size increases.
    
First, we present the system-size scaling results for the IPR, as shown in Fig.~\ref{fig:IPRplot} in the main text. This analysis quantifies the extent to which the wave function of the eigenstates of the Hamiltonian spreads in Hilbert space through the expression $-\ln\operatorname{IPR}_2\approx D_2\ln{N}$, where $N$ is the dimension of the symmetry-resolved Hilbert subspace. In other words, when the graph is fitted by a linear function, the slope corresponds to the quantity $D_2$ in the above formula, which is called the fractal dimension. Generally, in ergodic systems $D_2\approx 1$, whereas in MBL systems $D_2\approx 0$ (in practice, due to finite-size effects, $D_2\approx 0.1$). Our results are shown in Fig.~\ref{fig:fractaldim}. When fitting these data, in the case of $S=1$, we find $D_2\approx 1.04$ for $\mu=0.5$ (ergodic), and $D_2\approx 1.01$ for $\mu=3.0$, with no significant difference observed within the limits of our analysis. In contrast, for $S=1/2$, we obtain $D_2\approx 1.08$ for $\mu=0.5$ (ergodic), and $D_2\approx 0.76$ for $\mu=3.0$, indicating that when quasiperiodicity of the interactions is strong, multifractal characteristics appear. Considering that, in the $S=1$ case, the system size $L$ is smaller than in the $S=1/2$ case and lies near the experimental limit at which quantum many-body properties can be observed, one may expect that increasing the system size for $S=1$ would also reveal nonergodic behavior. At the very least, this provides a strong motivation for experimental investigation of this system.

\begin{figure}[H]
    \centering
    \includegraphics[width=\linewidth]{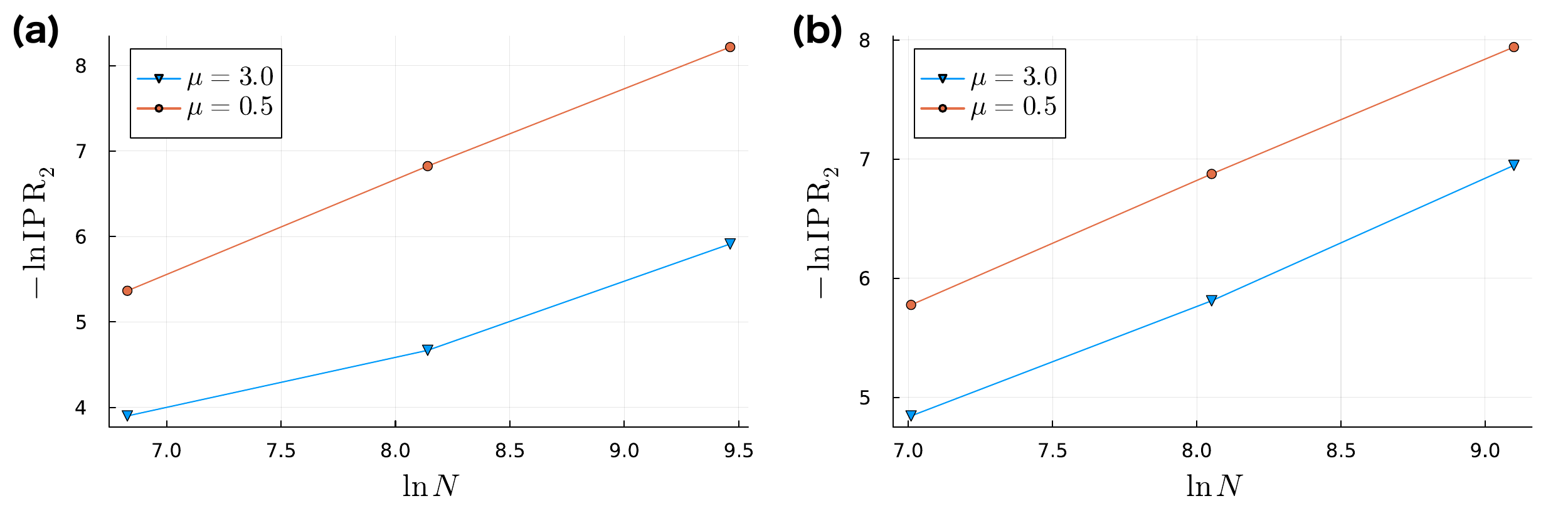}
    \caption{System-size scaling of the negative logarithm of the IPR. (a) Spin-1/2 systems with system sizes $L=12,14$, and $16$ in the $S^z_{\text{tot}}=0$ and $I=1$ sector. (b) Spin-1 systems with system sizes $L=8,9$, and $10$ in the $\tau^z_{\text{tot}}=0$ and $I=1$ sector. We take the average over 100 samples of $\delta\sim\mathcal{U}[0,2\pi]$ and over 100 states in the middle of the spectrum.}
    \label{fig:fractaldim}
\end{figure}

Second, we present the system-size scaling results for the bipartite EE. In general, when the partitioned subsystems have equal sizes, the bipartite EE in a thermal system satisfies $S\approx \frac{1}{2}\ln N$, where $N$ is the dimension of the total Hilbert space. In other words, when the graph is fitted by a linear function, the slope is $a=1/2$. Our results are shown in Fig.~\ref{fig:EE_scaling}. When fitting these data in the case of $S=1/2$, we find $a\approx 0.62$ for $\mu=0.5$ (ergodic), and $a\approx 0.46$ for $\mu=2.5$. Although in the case of $\mu=0.5$ the slope exceeds 0.5 significantly, this can be attributed to finite-size effects, in particular the underestimation of the EE for $L=12$. On the other hand, in the case of $S=1$, we obtain $a\approx 0.57$ for $\mu=0.5$, and $a\approx 0.48$ for $\mu=2.5$, showing a similar trend. For both $S=1/2$ and $S=1$, we could not confirm a clear deviation of the scaling coefficient $a$ from the thermal value at large $\mu$, although $a$ at large $\mu$ is smaller than that at small $\mu$. Therefore, the entanglement-entropy scaling analysis alone does not provide a decisive diagnostic for distinguishing the thermal regime from the possible many-body critical regime; rather, the smaller fitted coefficient at large $\mu$ should be regarded as a supporting observation.

\begin{figure}[H]
    \centering
    \includegraphics[width=\linewidth]{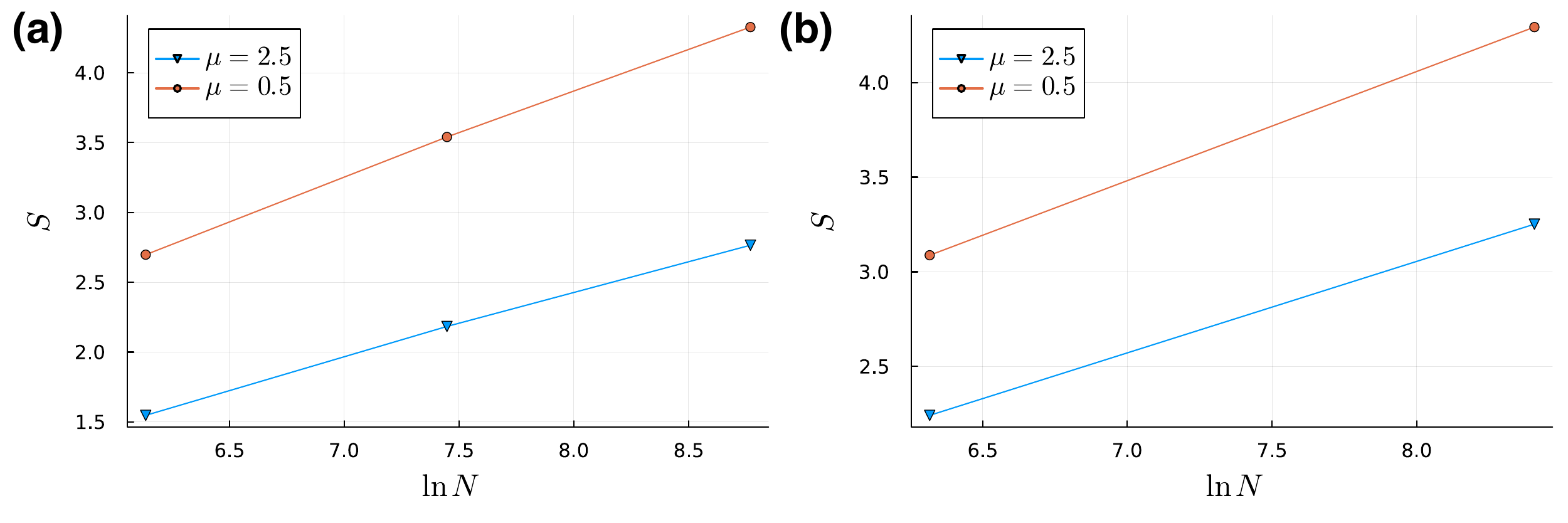}
    \caption{System-size scaling of the bipartite entanglement entropy. (a) Spin-1/2 systems with system sizes $L=12,14$, and $16$ in the $S^z_{\text{tot}}=0$ and $I=1$ sector. (b) Spin-1 systems with system sizes $L=8$ and $10$ in the $\tau^z_{\text{tot}}=0$ and $I=1$ sector.  We take the average over 100 samples of $\delta\sim\mathcal{U}[0,2\pi]$ and over 100 states in the middle of the spectrum.}
    \label{fig:EE_scaling}
\end{figure}

\section{Effects of long-range interactions and deviations from the ideal spin-1 $XY$ model}\label{appE}

In this appendix we investigate the effects of long-range interactions and deviations from the ideal condition $J_1=J_2=J_3$, which were neglected in the main text as approximations but are inevitable in actual experiments. To assess the effects of the approximations, we calculate the IPR for $S=1$ systems with $L=8, 9$, and $10$ in two situations: one including interactions up to fourth-nearest-neighbor terms and the other incorporating the realistic dipole-dipole interaction given by Eq.~(\ref{IntHamA4}), which arises due to the deviation from the ideal condition $J_1=J_2=J_3$. The results are shown in Fig.~\ref{fig:IPR_benchmark}. In the first situation, shown in Fig.~\ref{fig:IPR_benchmark}, $-\ln {\rm IPR}_2$ deviates from the thermal value around $\mu=1.0$. In the second situation, shown Fig.~\ref{fig:IPR_benchmark}, the growth of $-\ln{\rm IPR}_2$ toward the thermal value is suppressed in the large-$\mu$ regime, in contrast to the results in the main text. These results indicate that the tendency toward nonergodic behavior is not an artifact of truncating the interaction to the next-nearest-neighbor terms or of assuming the ideal condition $J_1=J_2=J_3$.

\begin{figure}[H]
    \centering
    \includegraphics[width=\linewidth]{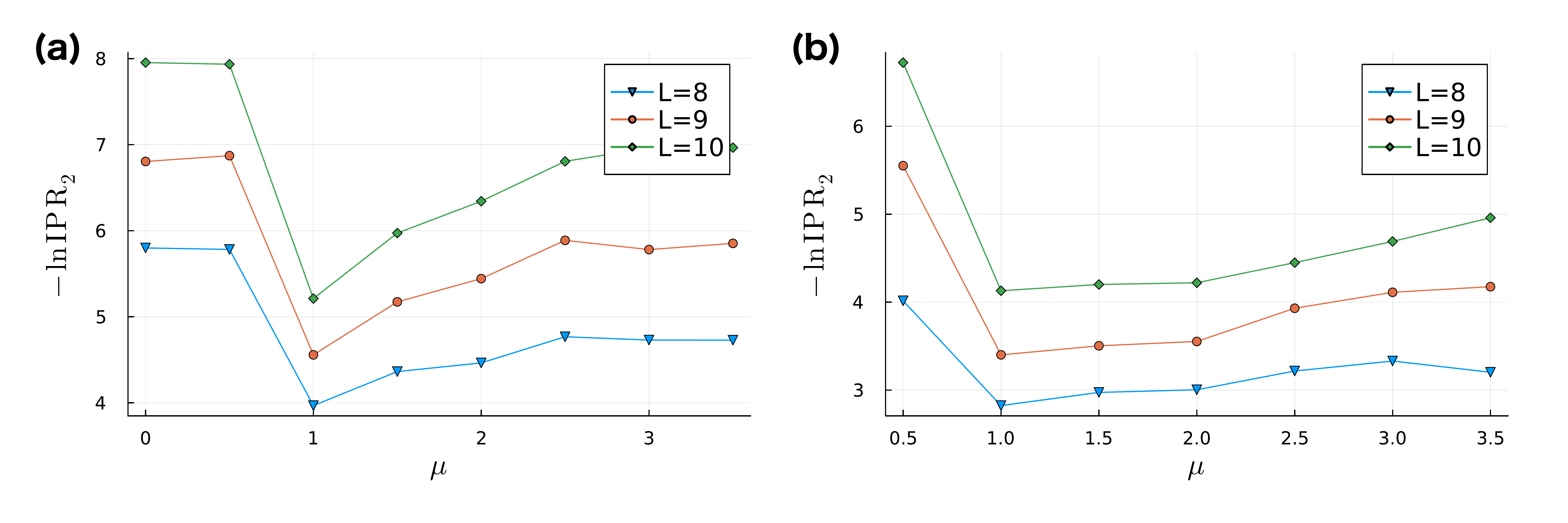}
    \caption{Negative logarithm of the IPR for $S=1$ systems with sizes $L=8,  9$, and $10$ for (a) a Hamiltonian with longer-range interactions (up to fourth-nearest-neighbor terms), and (b) a Hamiltonian with the realistic dipole-dipole interaction shown in Eq.~(\ref{IntHamA4}) (up to second-nearest-neighbor interactions). In both cases, we take the average over 50 samples of $\delta\sim \mathrm{Uniform}[0,2\pi]$ and 100 eigenstates in the middle of the spectrum.}
    \label{fig:IPR_benchmark}
\end{figure}

\end{widetext}


\begin{thebibliography}{73}
\bibitem [{foo()}]{footnote:interaction}%
  \BibitemOpen
  \href@noop {} {}\bibinfo {note} {We evaluated the average amplitudes of the first-, second-, and third-nearest-neighbor interactions for a system of size $L=16$. We define $\bar{J}_{r} \equiv \overline{|J_{i,i+r}|}$ $(r=1,2,3)$, where the bar denotes an average over sites $i$ and realizations of $\delta$. We find that $\bar{J}_{1}/J_0 \simeq 1.75$, $\bar{J}_{2}/J_0 \simeq 0.12$, and $\bar{J}_{3}/J_0 \simeq 0.01$ for $\mu=2.5$, and $\bar{J}_{1}/J_0 \simeq 1.01$, $\bar{J}_{2}/J_0 \simeq 0.13$, and $\bar{J}_{3}/J_0 \simeq 0.04$ for $\mu=0.5$.}\BibitemShut {Stop}%
\end{thebibliography}

%

\end{document}